\documentstyle[preprint,aps,floats]{revtex}
\begin{document}
\preprint{\tighten \vbox{\hbox{UT-preprint-0106}
  \hbox{hep-th/0111109}}}
\draft
\title{On Gauge Invariance of 
Noncommutative Chern-Simons Theories
}
\author{Guang-Hong Chen and 
Yong-Shi Wu}
\address{Department of Physics, 
University of Utah,\\
Salt Lake City, Utah  84112\\
\vspace{.5cm} 
{\tt ghchen@physics.utah.edu \\
     wu@physics.utah.edu}
}

\maketitle
\begin{abstract}
\baselineskip=0.90cm

Motivated by possible applications to condensed 
matter systems, in this paper we construct $U(N)$ 
noncommutative Chern-Simons (NCCS) action for a 
disc and for a double-layer geometry, respectively. 
In both cases, gauge invariance severely constrains 
the form of the NCCS action. In the first case, 
it is necessary to introduce a group-valued boson 
field with a non-local chiral boundary action, 
whose gauge variation cancels that of the bulk action. 
In the second case, the coefficient matrix $K$ 
in the double $U(N)$ NCCS action is restricted to be 
of the form  $K=k\left(\begin{array}{cc}1 
& 1 \\ 1 & 1 \end{array}\right)$ with integer 
$k$. We suggest that this double NCCS theory with 
$U(1)$ gauge group describes the so-called Halperin 
$(kkk)$ state in a double-layer quantum Hall system. 
Possible physical consequences are addressed.  

\end{abstract}
\newpage

\section{Introduction}

In the past two decades, Chern-Simons (CS) field 
theory\cite{earlycs,deser-rao,witten88,ZHK,frank} 
in $2+1$ dimensions has received a great deal of 
attention in theoretical physics. In particular,
when a $U(1)$ CS field couples to a matter field, 
statistics of the latter gets transmuted: The CS 
coupling automatically produces a CS flux associated 
with the matter particle. The composite of a particle 
and CS flux is called an anyon, whose exchange 
statistics depends on the CS coupling constant. The 
feat to transmute the exchange statistics in two 
spatial dimensions turned out to be extremely helpful 
in understanding the fractional quantum Hall effect  
and in the early attempt of anyon superconductivity
\cite{ZHK,frank}. It is also amusing to note that 
pure CS field theory itself can serve as the low 
energy effective theory for the quantum Hall 
systems \cite{wen-zee,wenzee2}. In this formulation, 
gauge invariance of the CS field theory plays an 
important role. For a finite system with an edge, 
Wen\cite{wen-zee} pointed out that a boundary term 
has to be added to maintain gauge invariance of 
the system, leading to his chiral Luttinger liquid 
theory of the quantum Hall edge states. Wen and
Zee, and Fr\"olich and Zee\cite{wenzee2} also 
systematically studied the structures of topological 
fluids in the framework of a multiple $U(1)$ CS theory.  

Recently field theories (especially gauge field 
theory) on a noncommutative space have attracted 
much interest \cite{douglas}.  Originally, such 
theory arises from string/M(atrix) 
theory \cite{matrix,witten}, which suggests 
that space or spacetime noncommutativity should 
be a general feature of quantum gravity for generic 
points deep inside the moduli space of M-theory. 
However, the simplest example of noncommutative 
space appears in a familiar quantum mechanical 
problem; namely, a charge in the lowest Landau level 
(LLL) in a strong magnetic field can be viewed as 
living in a noncommutative space, because the 
guiding-center coordinates of the charge are known 
not to commute\footnote{For a pedagogical review, 
see Ref. \cite{jackiw}.}. This fact motivates many 
(include the present authors) to believe that perhaps 
the most natural field theory formulation for the 
quantum Hall systems should be a noncommutative one. 
Indeed, along this route, Susskind \cite{susskind} 
recently argued that $U(1)$ noncommutative 
Chern-Simons (NCCS) theory at level $k$ on a plane 
is equivalent to the Laughlin theory at filling 
factor $\nu=1/k$. This scenario can be considered 
as a projection (or deformation) of the ordinary CS 
effective field theory formulation \cite{wenzee2} 
of quantum Hall fluids to the LLL which is 
intrinsically a noncommutative space. For further 
studies on Susskind's proposal see, e.g., Refs. 
\cite{simeon,poly,hansson}. 

In an ordinary non-Abelian CS theory the CS coupling 
(or the level) acquires, due to self interactions, 
a one-loop shift by the integer $N$ for $SU(N)$ 
($N\ge 2$) gauge group. This perturbative result 
provides an important test for the level quantization 
and other topological features of non-Abelian CS theory 
\cite{earlywu}. Since a $U(1)$ NCCS theory is no longer 
a free theory, it is natural to ask whether the self
interactions arising from noncommutativity will also
induce a similar one-loop shift in the $U(1)$ NCCS 
theory. This was confirmed in Ref. \cite{chen-wu} by 
an explicit one-loop calculation: Indeed, the CS 
coupling (or the level) is {\em shifted by one for 
$U(1)$ and by $N$ for $U(N)$ NCCS theory}. This was, 
to our knowledge, the first evidence for the level 
quantization in $U(1)$ NCCS theory, which was later 
discussed by other groups \cite{bak,nair} using 
topological arguments. Due to this quantum shift, 
Susskind's level-$k$ NCCS theory should describe 
a $\nu=\frac{1}{k+1}$ (rather than $\frac{1}{k}$) 
Laughlin state. This point has also been emphasized 
in recent papers\cite{poly,suss-hell}.

Motivated by the possibility that pure NCCS theory 
provides a low energy effective theory of quantum 
Hall fluids, in this paper we study how to construct 
NCCS theory for spatial geometry other than an 
infinite plane, that corresponds to more realistic
geometry for samples used in experimental studies
\cite{das}. The first case we will study is a sample 
of finite size with an edge, say a disc. The second 
case involves two spatially close but separated 
quantum Hall fluids in a double quantum well structure.
The two layers are coupled to form an interesting 
system, called a double-layer quantum Hall system, 
either due to tunneling or due to Coulomb interactions 
between the layers. We will mainly focus on how to 
construct a {\it gauge invariant} NCCS action for 
these sample configurations.

In the rest of this section, we establish our 
notations and conventions. A $(2+1)$-dimensional 
noncommutative spacetime has coordinates satisfying 
\begin{equation}
 [x^\mu, x^\nu]=i\theta^{\mu\nu} 
\hspace{1cm} \mu, \nu = 0,1,2, 
\label{noncom}
\end{equation}
where $\theta^{\mu\nu}$ are anti-symmetric and real 
parameters of dimension length squared. To realize 
the commutators (\ref{noncom}), we follow Moyal's
deformation techniques, adopting a representation 
in which the coordinates $x^\mu$ are commuting as 
usual, but the product of any two functions of 
$x^\mu$ is deformed to the star product:
\begin{equation} 
\label{starproduct}
(f*g)(x)=\exp\biggl[ {\frac{i}{2}\theta^{\mu\nu}
\partial^{f}_{\mu}\partial^{g}_{\nu}}\biggr] f(x)g(x).
\label{star}
\end{equation}
Then the commutators (\ref{noncom}) are interpreted 
as the bracket with respect to the star product:
\begin{equation}
 [x^{\mu},x^{\nu}]\equiv 
x^{\mu}*x^{\nu}-x^{\nu}*x^{\mu}.
\end{equation}
The advantage of this representation is that we can 
use the usual differentiation and integration, and 
in particular we can {\it define the boundary of 
noncommutative space or spacetime in the usual way}. 
Of course, in this case we have to pay attention to 
the non-locality arising in the theory on the boundary 
due to coordinate noncommutativity in the bulk. For 
applications to a system in the LLL, one considers 

only the spatial noncommutativity:
$\theta^{01}=\theta^{02}=0$ and $[x^{1},x^{2}] =i\theta$.

The action for a pure $U(N)$ CS theory on this 
spacetime reads 
\begin{eqnarray}
\label{nccs}
I_{CS}&=&-\frac{ik}{4\pi}\int_{{\cal M}} d^3x
\varepsilon^{\mu\nu\lambda}\mbox{Tr}(
A_{\mu}*\partial_{\nu}A_{\lambda}
+\frac{2}{3}A_{\mu}*A_{\nu}*A_{\lambda}), \\
\nonumber
&=&-\frac{ik}{4\pi}\int_{{\cal M}}\mbox{Tr}
(AdA+\frac{2}{3}A^3).
\end{eqnarray}
Here the dynamical fields are the gauge 
potential $A^{\mu} = A^a_\mu T^a$, with 
$T^a$ the generators of the gauge group 
$G=U(N)$, normalized to $ \mbox{Tr}(T^a T^b)
= - \delta^{ab}/2$, with $T^{0}=i/\sqrt{2N}$ 
for the $U(1)$ sector. Moreover, $k$ is the 
CS coupling, or the level parameter, and
$\varepsilon^{\mu\nu\lambda}$ the totally 
anti-symmetric tensor with $\varepsilon^{012}=1$. 
The integral is taken over a three dimensional 
spacetime ${\cal M}$, which will be specified 
later. In the second line of eq. (\ref{nccs}), 
we introduced the gauge potential one-form 
$A=A_{\mu}dx^{\mu}$. We also {\it suppress both 
the star product symbol and the wedge product 
symbol} in the forms. They are understood as 
\begin{equation}
\label{symbol}
A^3=A\wedge A\wedge A=A_{\mu}*A_{\nu}*A_{\rho}
dx^{\mu}\wedge dx^{\nu}\wedge dx^{\rho}.
\end{equation}
In our convention, the field strength
two-form is given by $F=dA+A^2$.

It is well-known that in the LLL the transverse 
plane, where the electrons live, becomes 
noncommutative. However, this does not forbid 
us to think that in a real sample of Hall bar or 
corbino geometry, there are electronic states that 
live on the edge of the sample. We emphasize that 
our adoption of the Moyal representation is 
crucial for providing a legitimate definition 
of a noncommutative space with edge or boundary. 
Indeed, in this representation, though the product 
in the function algebra is deformed, coordinates 
are still the ordinary ones, so the boundary of a 
space with finite geometry can be easily identified 
and dealt with as usual. (If we had realized the 
coordinates as operators, it would be hard to 
identify the boundary of a noncommutative space). 
Of course, we have to be careful about the product 
of two functions on the boundary which becomes 
non-local due to noncommutativity in the bulk.

The paper is organized as follows. In the Section 
II, we discuss the noncommutative edge theory, 
described by the boundary terms in the complete 
action of a $U(1)$ NCCS theory, for a quantum Hall 
fluid of disc geometry. In Section III, we discuss 
the multiple $U(1)$ NCCS theory for a double-layer 
quantum Hall system. In both cases the central issue 
is gauge invariance which, as we will see, severely 
constrains the form of the NCCS actions. Finally, 
conclusions and remarks are given in the Section IV.

\section{Edge theory of a quantum Hall fluid }

In this section, we study the NCCS on a  
cylindrical spacetime ${\cal M}=D\times R$, 
where $D$ stands for a disc, $R$ the time 
line. Though the main applications would be
the case with $U(1)$ gauge group, it is not 
hard to first present the more general case 
with $U(N)$ gauge group. Under a gauge 
transformation given by the map $g$, 
$g: {\cal M}\to U(N)$, the gauge potential 
$A$ and field strength $F$ change to 
\begin{equation}
\label{Aprime}
A^{g}=gAg^{-1}+gdg^{-1};\hspace{1.0cm} 
F^{g}=gFg^{-1}.
\end{equation}
Note that the star product is understood hereafter
and $g^{-1}$ is defined by $g*g^{-1}=g^{-1}*g=1$. 
Correspondingly, the NCCS action (\ref{nccs}) 
transforms into
\begin{equation}
\label{trannccs}
I_{cs}(A^{g})=I_{cs}(A)+\frac{ik}{4\pi}
\int_{D\times R}\mbox{Tr}d(dg^{-1}gA)+\frac{ik}
{12\pi}\int_{D\times R}\mbox{Tr}(gdg^{-1})^3.
\end{equation}
Usually ${\cal M}$ is assumed to have no boundary,
then the second term vanishes. The third (the 
Wess-Zumino-Witten) term in eq. (\ref{trannccs}) 
would be a topological term, which does not vanish 
for a topologically nontrivial (or large) gauge 
transformation. Gauge invariance of the path 
integral measure in the NCCS theory then imposes 
the level quantization condition, namely the 
level $k$ has to be an integer; so that the 
partition function is invariant under a large 
gauge transformation. This topological argument, 
which restricts the quantum one-loop shift in
$k$ has to be integer too, holds true in the 
noncommutative case even for the $U(1)$ case
\cite{chen-wu,bak,nair}. Here we see the pivotal
importance of the fundamental requirement of 
gauge invariance in CS theories, ordinary or 
noncommutative.


However, in our present case the cylindrical 
spacetime ${\cal M}=D\times R$ is not without
boundary. Under an infinitesimal gauge 
transformation, $g=1+\varepsilon$, we have
\begin{equation}
\label{infinitesimal}
\delta_{\varepsilon} A=A^{g}-A=
-d\varepsilon-[A,\varepsilon]=-D\varepsilon.
\end{equation}
By using the Bianchi identity
\begin{equation}
\label{bianchi}
DF=dF+[A,F]\equiv 0,
\end{equation}
we get the gauge variation of the action 
(\ref{nccs}) as 
\begin{equation}
\label{deltaI}
\delta_{\varepsilon}I_{cs}=\frac{ik}{4\pi}
\int_{D\times R}\mbox{Tr}d(\varepsilon dA)
=\frac{ik}{4\pi}\int_{D\times R}
\mbox{Tr} d(\varepsilon(F-A^2)).
\end{equation}
It is obvious now that the NCCS action on 
$D\times R$ is not gauge invariant any more. 

Since any microscopic theory of the quantum 
Hall effect should be gauge invariant, the 
gauge non-invariance of the NCCS action 
(\ref{nccs}) on $D\times R$ tells us that 
{\em a pure NCCS action in the bulk is not 
a complete effective description for quantum 
Hall fluids on a finite geometry with edge}. 
Since the gauge anomaly term (\ref{deltaI}) 
is a total divergence and, therefore, can be 
written as a boundary term, there should be 
a way to cancel it by adding ``boundary" 
terms. This is the essence of the 
Callan-Harvey effect \cite{callan-harvey}, 
where gauge anomalies are canceled between 
two systems with different dimensionality.  

It is known that the NCCS theory is not 
invariant under time reversal \cite{jabarri}. 
This is just right, since the microscopic 
theory of the quantum Hall system involves 
charged particles in a magnetic field, which 
explicitly breaks the time reversal symmetry. 
Now that the motion of a quantum Hall fluid 
is intrinsically chiral, it is natural to look 
for a chiral theory on the edge to cancel the 
gauge anomaly (\ref{deltaI}). Here we follow 
the procedure in Ref. \cite{stone}, to look for 
a deformed version of a chiral boson action on 
the edge which, when combined with the bulk 
NCCS action, makes the total action gauge 
invariant. With the experience of bosonizing 
a chiral fermion system, we propose that the 
following is the desired edge action for a 
noncommutative chiral boson $h(x,t,\tau) \in G$:
\begin{eqnarray}
\label{s0}
S_0&=&\frac{ik}{4\pi}\int_{D\times R}dxdtd\tau 
\mbox{Tr}\partial_{\tau}(h^{-1}*
\partial_{x}h*h^{-1}*\partial_x{h})\\ \nonumber
&+&\frac{ik}{2\pi}\int_{D\times R}dxdtd\tau
\mbox{Tr}(h^{-1}*\partial_{t}h*
\partial_x(h^{-1}*\partial_{\tau}h)).
\end{eqnarray}
Here $x$ and $\tau$ are the angular and radial 
coordinates on the disc $D$, while $t$ the time 
coordinate on $R$. (See Fig. 1.) 

To show such an action is indeed chiral, an 
elegant way \cite{stone2} is to connect this 
action to the noncommutative Wess-Zumino (NCWZ) 
action in three dimensions: 
\begin{equation}
\label{WZ}
S_{wz}=\frac{1}{12\pi}\int_{D\times R}
\mbox{Tr}(h^{-1}dh)^3.
\end{equation}
In fact, the above Wess-Zumino term can be 
rewritten as
\begin{equation}
\label{wz2}
S_{wz}=\frac{1}{4\pi}\int_{D\times R}dxdtd\tau
\mbox{Tr}([h^{-1}*\partial_xh,h^{-1}*
\partial_{t}h]*h^{-1}*\partial_{\tau}h),
\end{equation}
after several integrations by parts and 
using the identity
\begin{equation}
\label{identity}
\partial_{\mu}(h^{-1}*\partial_{\nu}h)
-\partial_{\nu}(h^{-1}*\partial_{\mu}h)+[h^{-1}*
\partial_{\mu}h,h^{-1}*\partial_{\nu}h]=0,
\end{equation}
which is nothing but the flat connection 
conditions for the one-form 
$h^{-1}dh$. Finally we write 
the NCWZ term (\ref{WZ}) as
\begin{eqnarray}
\label{connection}
S_{wz}&=&\frac{1}{2\pi}\int_{D\times R}dxdtd\tau
\mbox{Tr}[(h^{-1}*\partial_th)*\partial_x(h^{-1}*
\partial_{\tau}h)] \\ \nonumber
&-&
\frac{1}{4\pi}\int_{D\times R}dxdtd\tau
\mbox{Tr}\partial_{\tau}(h^{-1}*\partial_xh*
g^{-1}*\partial_th).
\end{eqnarray}

Thus we can rewrite the action (\ref{s0}) in 
terms of the NCWZ term as
\begin{equation}
\label{s02}
S_0=\frac{ik}{4\pi}\int_{D\times R}dxdtd\tau
\mbox{Tr}\partial_{\tau}[h^{-1}*\partial_xh
*h^{-1}*(\partial_x+\partial_t)h]+\frac{ik}
{12\pi}\int_{D\times R}\mbox{Tr}(h^{-1}dh)^3.
\end{equation}
The variation of this action (\ref{s0}) with
respect to $h$ gives
\begin{eqnarray}
\label{variationofs0}
\delta S_0&=&-\frac{ik}{2\pi}\int_{D\times R}
dxdtd\tau\mbox{Tr}
\partial_{\tau}[h^{-1}*\delta h*\partial_x
(h^{-1}*\partial_{+}h)]\\ \nonumber
&=&-\frac{ik}{2\pi}\int_{D\times R}dxdtd\tau
\mbox{Tr}\partial_{\tau}[\delta h*h^{-1}*
\partial_{+}(\partial_xh*h^{-1})].
\end{eqnarray}
Here $\partial_{+}=\partial_x+\partial_t$. 
Thus the equation of motion is given by
\begin{equation}
\label{equationofmotion}
\partial_{+}(\partial_{x}h*h^{-1})=0.
\end{equation}
(The exact meaning of the star product in 
this equation in 1+1 dimensions will be
clarified later in this section.) The 
solution of this equation of motion 
represents right-going waves. In this way, 
the action $S_0$ is the chiral version of 
the noncommutative deformation of the 
Wess-Zumino-Witten model.

Before adding the action $S_0$ to the bulk 
NCCS action (\ref{nccs}), we need to couple 
the NC chiral boson to the gauge fields
$A_{\mu}$, so that the resulting theory is 
gauge invariant under the combined variations
\begin{equation}
\label{combine}
\delta h=\varepsilon*h, 
\hspace{1.0cm}A\rightarrow A+
\delta_{\varepsilon}A.
\end{equation}
Recalling that the resulting theory has to 
be chiral, we couple the noncommutative 
current $\partial_xh*h^{-1}$ to the chiral 
combination of potentials $A_{+}=A_x+A_t$. 
After some calculations, one can check that 
under the combined transformation 
(\ref{combine}), the following action
\begin{equation}
\label{sga}
S(h, A)=S_0+\frac{ik}{2\pi}\int_{D\times R}
dxdtd\tau \mbox{Tr}\partial_{\tau}
[A_{+}*\partial_xh*h^{-1}]+\frac{ik}{4\pi}
\int_{D\times R}dxdtd\tau\mbox{Tr}
\partial_{\tau}(A_{+}*A_x), 
\end{equation}
has the desired variation 
\begin{eqnarray}
\label{resulting}
\delta_{\varepsilon}S(h,A)
&=&-\frac{ik}{4\pi}\int_{D\times R}dxdtd\tau
\mbox{Tr}\partial_{\tau}(\varepsilon dA), 
\\ \nonumber
&=&-\frac{ik}{4\pi}\int_{D\times R}
\mbox{Tr}d(\varepsilon dA),
\end{eqnarray}
which is just the opposite of the gauge 
variation (\ref{deltaI}) of the bulk NCCS 
action (\ref{nccs}). Therefore, the combined 
action 
\begin{equation}
\label{Stotal}
S_{total}=I_{cs}+S(h,A)
\end{equation}
is gauge invariant under the combined 
transformation (\ref{combine}), and thus 
serves as the complete effective action 
for a quantum Hall fluid on a disc. The 
gauge anomaly from the bulk NCCS theory 
is cancelled by an edge theory described 
by $S(h,A)$.

Before ending this section, we need to clarify 
more precisely the meaning of the star product 
in eq. (\ref{equationofmotion}) on the edge. 
Note that originally the star product is defined 
in the bulk of the disc. In eq. 
(\ref{equationofmotion}) we first do the star 
product as if in the bulk and then restrict 
the value of $\tau$ to be that on the boundary: 
i.e. $\tau=R$ with $R$ the radius of the disc. 
We observe that $(\tau, x)$ are actually the 
polar coordinates on the disc, while originally 
we have $[x^{1},x^{2}] =i\theta$ in terms of 
Cartesian coordinates. So the restriction of
eq. (\ref{star}) leads to the following 
explicit expression on the boundary
\begin{eqnarray}
(f*g)(x,t)= \exp \{\frac{i}{2}\theta 
[\cos x (\frac{\partial}{\partial \tau})_f
- \frac{\sin x}{\tau}(\frac{\partial}{\partial x})_f] 
[\sin x (\frac {\partial}{\partial \tau})_g +
\frac{\cos x}{\tau}(\frac {\partial}
{\partial x})_g]\nonumber\\  
- (f \leftrightarrow g)\} 
f(\tau,x,t)g(\tau,x,t)|_{\tau=R=1}.
\label{star2}
\end{eqnarray} 
Where we have set the radius of disc as unity.

In ordinary effective CS theory, the chiral 
boson field $h$ lives only on the edge. 
However in the NCCS theory the star product 
on the edge is inherited from that in the bulk, 
so the chiral boson field $h$ must be extended 
to the bulk. Thus, the edge action (\ref{sga}), 
though consisting of integrals on the boundary, 
actually depends on the extension of the chiral 
boson field $h$ from the edge to the bulk, 
because the star product (\ref{star2}) involves 
infinitely many higher order derivatives in the 
radial direction. So rigorously speaking, the 
chiral boson field $h$ is not restricted to 
the edge only. But its bulk action, the NCWZ 
term in the action (\ref{s02}), like its 
commutative counterpart, is a topological action
\cite{bak,nair}. Hence its dynamics is described only 
by the boundary terms in the actions (\ref{s02}) and 
(\ref{sga}). It is in this (weaker) sense that we 
say the additional action (\ref{sga}) describes 
a dynamical theory on the edge. We note that this 
agrees with the present physical picture in 
condensed matter community about the edge states 
in the quantum Hall effect: Namely, the edge 
excitations described by the above chiral boson 
are actually quasiparticles, which do exist in 
the bulk as gapful excitations, but become gapless 
and dynamical at low energy while on the edge.

\section{Double-Layer quantum Hall system}

In this section, we are going to construct 
a gauge invariant action for a quantum Hall 
double-layer system using NCCS effective field 
theory. We first briefly review Wen and Zee's 
formulation\cite{wen-zee}, in which a 
double-layer quantum Hall system is described
by the following Lagrangian for two ordinary 
$U(1)$ CS fields, one for each layer:
\begin{equation}
\label{WZLAG}
S=\frac{1}{4\pi}\int \sum_{I,J}\,
K_{IJ}\, A^I d\,A^J,
\end{equation}
where $I,J =1,2$ are layer indices. If the 
symmetric $K$-matrix does not have zero 
eigenvalue, then it describes an incompressible 
quantum Hall fluid, with the total filling factor
$\nu=\sum_{I,J}(K)^{-1}_{IJ}$. On the other 
hand, if the matrix $K$ has a zero eigenvalue, 
then some linear combination of the gauge 
fields becomes massless and the Hall fluid 
is compressible. In particular, when all the 
matrix elements take the same value, say $k$, 
such effective theory describes a system with
the total filling factor $\nu=\frac{1}{k}$
\cite{das,wen-zee2}. It is easy to check that 
the Lagrangian (\ref{WZLAG}) is gauge invariant 
on a compact space under the $U(1)\times U(1)$
gauge transformations
\begin{equation}
\label{comb-gauge}
A^{I} \rightarrow A^{\prime\,I}
=A^{I}+ d\,\lambda^{I},
\end{equation}
where $I=1,2$. Namely, the fields at different 
layers transform independently. In this 
formulation, gauge invariance of the CS theory 
does not require the symmetric matrix $K$ to 
be quantized. However, as pointed out by Wen and 
Zee, the sources of gauge potential $A_I$ 
are vortex-like quasi-particles in the fluid; it 
is their circulation quantization that fixes the 
elements of the $K$ matrix to be integers. 

Now we proceed to generalizing Wen-Zee's Lagrangian
(\ref{WZLAG}) for multiple $U(1)$ CS fields to the 
noncommutative case. Some new features are noted 
immediately. First, instead of eq. (\ref{comb-gauge}), 
the noncommutative $U(1)$ gauge transformation law 
in an individual layer is modified to  
\begin{equation}
\label{sepa-gauge}
\delta A^{I}= d \lambda^{I} + [A^I, \lambda^I], 
\end{equation}
(no summation over $I$). Therefore, noncommutative
gauge invariance requires that the $U(1)$ NCCS action 
for individual layers, i.e. the terms with $I=J$ in 
the action (\ref{WZLAG}), be supplemented by cubic 
terms, as those shown in eq.(\ref{nccs}). For the 
sake of generality, it is natural to introduce also 
new cubic cross-terms involving both $A^1$ and $A^2$, 
whose coefficients are to be determined by gauge 
invariance. These considerations lead to the 
following suggestion for the most general form of
the NCCS action for a double-layer system:
\begin{eqnarray}
\label{generic1}
S_{double} &=& K_{11}\int(A^{1}dA^{1}
+\frac{2}{3}A^{1}A^{1}A^{1})+K_{22}\int(A^{2}dA^{2}
+\frac{2}{3}A^{2}A^{2}A^{2}) \\ \nonumber
&+&K_{12}\int(A^{1}dA^{2}+L_{122}A^{1}A^{2}A^{2}) 
+ K_{21}\int(A^{2}dA^{1}+L_{211}A^{2}A^{1}A^{1}).
\end{eqnarray}
Obviously, if $\theta=0$, all the cubic terms 
automatically vanish and we return to the action 
(\ref{WZLAG}).



To examine gauge invariance of the action 
(\ref{generic1}), we first write the variation 
of the action (\ref{generic1}), under generic 
$\delta A^{1}$ and $\delta A^{2}$, as a sum 
of four terms:  
\begin{eqnarray}
\label{gaugevariation}
\delta S^{11}_{double}&=&2K_{11}
\int\delta A^{1}F^{1};\hspace{1.0cm}
\delta S^{22}_{double}
=2K_{22}\int\delta A^{2}F^{2}, \\ 
\delta S^{12}_{double}
&=&K_{12}\int\delta A^{1}(dA^{2}+L_{122}A^{2}A^{2})
+K_{12}\int\delta A^{2}[dA^{1}
+L_{122}(A^{2}A^{1}+A^{1}A^{2})], \\
\delta S^{21}_{double}
&=&K_{21}\int\delta A^{2}(dA^{1}+L_{211}A^{1}A^{1})
+K_{21}\int\delta A^{1}[dA^{2}
+L_{211}(A^{1}A^{2}+A^{2}A^{1})].
\end{eqnarray}

Now let us assume that $A^1$ and $A^2$ transform 
independently, as given by eqs. (\ref{comb-gauge}) 
with $\lambda_1$ and $\lambda_2$ independent 
of each other. Then the gauge variation of the
diagonal terms $\delta S^{11}_{double}$ and 
$\delta S^{22}_{double}$ are separately zero, 
while that of the cross terms, 
$\delta S^{12}_{double}+ \delta S^{21}_{double}$ 
vanishes only if
\begin{equation}
K_{12}+K_{21}=0,\qquad L_{122}=L_{211}=0.
\label{trivial}
\end{equation}
Substituting these back to the action (\ref{generic1}), 
we see that the cross terms are identically zero. 
This leads to completely decoupled two layers, not 
interesting at all. 

In a double-layer quantum Hall system, the two 
layers can couple to each other either due to 
interlayer tunneling or due to Coulomb 
interactions between charged particles in different 
layers. This will naturally lead to the appearance 
of the cross terms in the effective theory. The 
result (\ref{trivial}) tells us that in order for 
an action like eq. (\ref{generic1}) to describe 
a {\it correlated} double-layer quantum Hall system, 
one can not assume the two $U(1)$ NCCS fields 
transform independently. This is an amazing consequence 
of gauge invariance in a multiple $U(1)$ NCCS theory. 

Then we ask: what kind of modification of the gauge 
transformations will be allowed that is compatible
with NCCS gauge invariance? We need cancellations
in the gauge variations of the cross terms. So this 
motivates us to propose that when the NCCS field in 
one layer transforms like a gauge potential, that in 
the other layer transforms covariantly like an
adjoint matter:  
\begin{equation}
\label{typeone}
\delta_{J}A^{I}=\delta^{I}_{J}\delta\lambda^{I}
+[A^{I},\lambda^{J}],
\end{equation}
where $I,J=1,2$ and no summation is assumed for 
repeated indices. First let us take $J=1$ in the 
combined gauge transformation (\ref{typeone}), then
\begin{equation}
\label{explicit}
\delta A^{1}=D\lambda^{1}, \hspace{1.5cm}
\delta A^{2}=[A^{2},\lambda^{1}].
\end{equation}
Consequently, the gauge variations in 
eq. (\ref{gaugevariation}) are evaluated as
\begin{eqnarray}
\label{gaugevariation2}
\delta S^{11}_{double}&=&0;\hspace{1.0cm}
\delta S^{22}_{double}=-2K_{22}\int 
d\lambda^{1}A^{2}A^{2}, \\ 
\delta S^{12}_{double}&=&K_{12}L_{122}
\int d\lambda^{1}A^{2}A^{2}+K_{12}\int\lambda^{1}
d(A^{2}A^{1}+A^{1}A^{2}), \\
\delta S^{21}_{double}&=&K_{21}(1-L_{211})
\int\lambda^{1}d(A^{1}A^{2}+A^{2}A^{1}).
\end{eqnarray}
Thus the gauge invariance, $\delta S_{double}=0$, 
requires that 
\begin{equation}
\label{condition1}
K_{12}L_{122}=2K_{22};\hspace{1.5cm}
K_{12}+K_{21}=K_{21}L_{211}.
\end{equation}
Similarly, we take $J=2$ and derive from gauge 
invariance the equations
\begin{equation}
\label{condition2}
K_{21}L_{211}=2K_{11};\hspace{1.5cm}
K_{12}+K_{21}=K_{12}L_{122}.
\end{equation}
Solving this two sets of equations, we get
\begin{eqnarray}
\label{solution}
K_{11}&=&K_{22}=\frac{K_{12}+K_{21}}{2}; 
\\ \nonumber
L_{122}&=&\frac{K_{12}+K_{21}}{K_{12}};
\hspace{1.0cm}L_{211}=\frac{K_{12}+K_{21}}{K_{21}}.
\end{eqnarray}
It is easy to show that we can always take 
the $K$-matrix to be symmetric, so it is 
constrained to be of the form
\begin{equation}
\label{Kmatrix}
K=k\left(\begin{array}{cc}1 
& 1 \\ 1 & 1 \end{array}\right). 
\end{equation}
Moreover, we have 
\begin{equation}
 L_{122}= L_{211}=2. 
\label{conseq}
\end{equation}
These are constraints required by the gauge 
invariance under the combined transformations 
(\ref{typeone}). 

In summary, we have obtained a gauge invariant 
double $U(1)$ NCCS action as follows:
\begin{eqnarray}
\label{generic}
S_{double} &=& \frac{k}{4\pi}\int(A^{1}dA^{1}
+\frac{2}{3}A^{1}A^{1}A^{1})+(A^{2}dA^{2}
+\frac{2}{3}A^{2}A^{2}A^{2}) \\ \nonumber
&+& \frac{k}{4\pi} \int(A^{1}dA^{2}+ 2A^{1}A^{2}A^{2}) 
+  (A^{2}dA^{1}+ 2 A^{2}A^{1}A^{1}).
\end{eqnarray}
The gauge transformation laws are those given in 
eqs. (\ref{typeone}). We suggest that this NCCS 
theory describes the Halperin $(kkk)$ state for 
a double-layer quantum Hall system.

Incidentally, we note that if we include in the 
integral sign the trace in group space as well, 
then the above result (\ref{generic}) applies to
the double U(N) NCCS theory too. If we further 
take the NC parameter $\theta=0$, then it reduces 
to a gauge-invariant action for coupling two 
single-layer non-Abelian Chern-Simons fluids, 
with $SU(N)$ gauge group, into a correlated 
double-layer system. To our knowledge, such a 
construction did not exist before even for 
ordinary CS theory.  

Returning to the double $U(1)$ NCCS action   
(\ref{generic}), one immediate question coming 
to our mind is whether the level $k$ should be
quantized or not? In the commutative counterpart, 
there is no topological argument for the 
quantization of $k$; but Wen and Zee \cite{wen-zee}
have argued that the elements of the $K$-matrix 
should be integers, based on the circulation 
quantization of the vortex-like quasiparticles 
in the theory. Here we conjecture that there 
should be a one-loop quantum shift of the level 
parameter $k$ in the double $U(1)$ NCCS theory, 
and its value should be unity, just like the 
single-layer $U(1)$ NCCS theory as computed by 
us in Ref. \cite{chen-wu}. The quantized one-loop 
shift of $k$ is perhaps constrained by a topological 
argument, which generalizes a similar argument in
the single-layer NCCS case\cite{bak,nair} 
and leads to the quantization of level $k$ as 
well.

\section{Conclusions and Discussions}

In this paper we have generalized the construction
of the $U(N)$ NCCS theory from a single-layer plane 
to a single-layer disc and to a double-layer plane.  
Gauge invariance is the main issue we addressed.
 
In the first case, it is necessary to introduce a 
group-valued bosonic field $h$ with chiral boundary 
terms, whose gauge variation cancels that of the 
bulk NCCS action. Mathematical steps of this 
construction proceed much the same way as in the 
commutative counterpart\cite{stone}. But some new 
distinct features appear due to noncommutativity  
in the bulk. Essentially now the boundary action 
becomes non-local both in the edge and in the
radial direction of the disc, in contrast to the
ordinary case where the boundary action describes
local dynamics on the edge. Since the boundary
terms depend on the extension of the group-valued 
$h$-field in the bulk, it can no longer be viewed 
as degrees of freedom that only live on the boundary, 
though its bulk action is still a topological one 
(the NCWZ term). Mathematically these two new 
features arise due to the restriction to the edge 
of the noncommutative star product in the bulk, which 
involves infinitely many higher order derivatives 
in both the angular and the radial directions. 

When the gauge group is $U(1)$, the construction
gives a noncommutative deformation of some known 
theory for edge states of quantum Hall (QH) fluids
(see Refs. \cite{wen-zee,stone}). It would be interesting 
to see what new physics the aforementioned non-local 
features of our construction would bring to the 
QH edge theory. Obviously the non-locality of
the boundary terms would make the interactions
along the edge and the bulk-edge connections more 
complicated and intriguing than we thought before.
Of course, one of the key questions is whether or 
not and, if yes, how the chiral Luttinger-liquid 
exponents could be affected by all these complications. 
An analysis based on renormalization group (RG) is 
needed to answer this question. Indeed an RG analysis 
for the noncommutative Landau-Ginsburg theory on a 
plane has been done already in Ref. \cite{chen-wu2}; 
however, the case at hand is quite different and 
presents several new features. We hope to have 
chance to return to this problem in the future.  

In the case of a double-layer system, gauge 
invariance does not constrain ordinary multiple 
$U(1)$ CS action, while the constraints in the
non-Abelian $U(N)$ CS case have never been worked 
out. In this paper we have found that for an NCCS 
theory on a {\it correlated} double-layer, gauge 
invariance indeed severely constrains the form
of the $K$-matrix in the action: Namely all
elements of the $K$-matrix have to  be the same 
integer. Experimentally, there is a state in
the double-layer quantum Hall system, the
Halperin $(kkk)$ state, which is believed to 
be described, at least, by the ordinary double 
$U(1)$ CS theory with a similar $K$-matrix.
It would be interesting to see whether our 
noncommutative deformation could be used to
describe this state as we have suggested. 
In particular, there are new (cubic) terms 
appearing in the cross action that mixes the 
CS fields at different layers, as required by 
gauge invariance. What would be the physical 
effects of these terms? We also note that if 
we take the NC parameter $\theta=0$, our result 
(\ref{generic}) gives a new action that 
generalizes multiple $U(1)$ ordinary CS theory 
to multiple {\it non-Abelian} $U(N)$.  

One question that puzzles us is the following:
Many of the $K$-matrix in double $U(1)$ 
ordinary CS theory that are discussed in the 
literature on the QH edge states are {\it not} 
of the very restrictive form (\ref{Kmatrix}) 
as derived in this paper. The question is
whether it is possible to get NC deformation
of CS theory with those $K$-matrices? On one
hand, we feel it should be possible. On the 
other hand, we do not know how to do it. The
point is that we have made a specific 
assumption (\ref{typeone}) on how one CS 
field transforms under the gauge transformation 
of the other. Could we relax this assumption 
to get the NC deformation of multiple CS 
theory with more general $K$-matrix? 
 
Finally let us mention that in relation to string 
theory, we think it could be interesting to work 
out the brane realization in the spirit of Ref. 
\cite{bergman,suss-hell}, or the CS matrix realization
in the spirit of Ref. \cite{susskind,poly}, of the NCCS 
theories we discussed above. 
 
\acknowledgments
This research was supported in part by the US National 
Science Foundation under Grants No. PHY-9970701.


\newpage
\begin{figure}
\caption{Disc Geometry}
\end{figure}
\begin{figure}
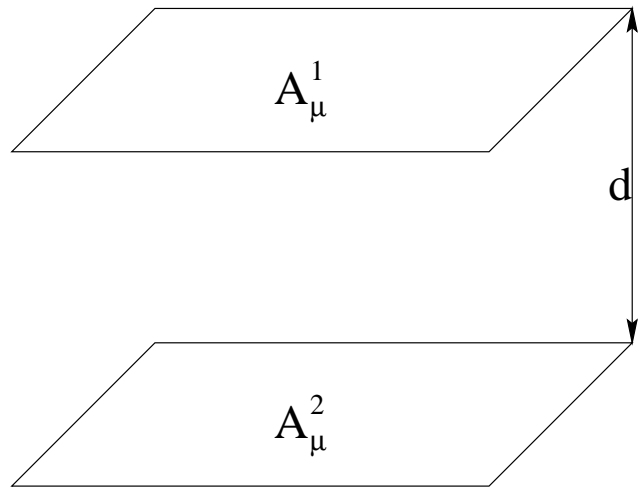

\caption{ Double-layer geometry }
\end {figure}

\end{document}